 \documentstyle[prl,aps]{revtex}  
\begin{document}
\draft


 \twocolumn[\hsize\textwidth\columnwidth\hsize  
 \csname @twocolumnfalse\endcsname              

\title{ Theoretical Constraints for Observation \\
of Superdeformed Bands in the Mass-60 Region\\
}
\author{
Yang Sun$^{(1)}$, Jing-ye Zhang$^{(1)}$, Mike Guidry$^{(1)}$,
and Cheng-Li Wu$^{(2)}$
}
\address{
$^{(1)}$Department of Physics and Astronomy, University of Tennessee,
Knoxville, Tennessee 37996 \\
$^{(2)}$Department of Physics, Chung Yuan Christian University, Chung-Li,
Taiwan 32023, ROC \\
}

\date{\today}
\maketitle

\begin{abstract}
The lightest superdeformed nuclei of the mass-60 region are described
using the Projected Shell Model.
In contrast to the heaviest superdeformed
nuclei where a coherent motion
of nucleons often dominates the physics,
it is found that alignment of
$g_{9/2}$ proton and neutron pairs
determines the high
spin behavior for superdeformed rotational bands
in this mass region.
It is predicted that, due to the systematics of
shell fillings along the even--even
Zn isotopic chain, observation of
a regular superdeformed yrast band sequence
will be unlikely for certain nuclei in this mass region.
\end{abstract}


 ]  

\narrowtext

The mass-190 nuclei are the heaviest nuclei
known where long-sequence rotational bands associated with the
superdeformed (SD)
minimum have been observed \cite{data_exp}.
In a recent
systematic study using the Projected Shell Model (PSM) \cite{psm},
it was concluded that the role of high-$j$
intruder orbitals is suppressed in these nuclei
because of strong correlations in the quadrupole field and non-negligible
correlations in the pair field \cite{sd190a}.
This conclusion was reinforced by the demonstration that quasiparticle
additivity generally does not hold \cite{zhang}.
Superdeformation in the mass-60 region
was predicted some years ago \cite{rag90} and was
recently observed \cite{zn62,cu58,zn60}.
This is the lightest known region of SD rotational bands  
and these new bands
show very different character from
those of the mass-190 nuclei. 

The mass-60 SD bands 
are associated with the highest rotational frequencies 
($\hbar\omega \approx$ 1.8 MeV)
observed so far in SD nuclear systems; in contrast, in the SD mass-190 nuclei
the maximum rotational frequency is typically 0.4 MeV.
However, the magnitudes of
deformation and
pairing appear to be comparable
in the mass-60 and  mass-190 regions.
We may expect that
the single-particle level density
near the $N$ = 30
gap is much lower than for heavier nuclei.
Thus, there can be substantial fluctuations in shell fillings
along an isotopic chain,
which could give rise to
drastic changes in the single particle and collective
behavior.
In addition, the 
maximum spin within the yrast and near-yrast bands in SD mass-60
nuclei is generally much
lower than in heavier nuclei (SD bands terminate earlier \cite{ring}).
These new features lead us to expect complex behavior in this region relative to
previously studied SD nuclei. 

So far, the SD bands of the mass-60 nuclei have been explained
using mean-field
theories
(cranked relativistic mean-field theory and cranked Nilsson model
\cite{ring}, or cranked Skyrme-Hartree-Fock method \cite{cu58}), with
complete neglect of pairing correlations.
These descriptions reproduce many of the
gross features found in these nuclei.
However, some interesting questions have not been
discussed. For example, why does the observed SD band in $^{62}$Zn \cite{zn62}
consist of only a few $\gamma$-rays, while population of
the SD band in neighboring $^{60}$Zn
\cite{zn60} extends to low spin states?
And why has one not seen a SD yrast band at all in $^{64}$Zn
\cite{zn64}?  
Can one predict spin values for these bands?
Can one give a
microscopic justification for the complete neglect of pairing in all
calculations reported prior to this one?  

In an investigation using the PSM, we have found
a surprisingly good description of the SD behavior
in this region and rather plausible answers to these questions
in terms of band crossings and band
interactions involving the $g_{9/2}$ intruder orbits.
Because of high angular momentum $j$, the fully paired $g_{9/2}$ 
quasiparticles in the ground state are most strongly affected by
the Coriolis Anti-pairing force when the nucleus rotates. 
The pairs break during the rotation 
and align their spins along
the direction of the collective rotation. Viewed in terms of bands, 
a 2-quasiparticle band (or a band with a broken pair)
which lies higher in energy at zero rotation 
becomes lower at a certain angular momentum than the ground band.
Thus, band crossing is related to the microscopic alignment process,
and can be linked to experimental observations. 
In this Letter, we concentrate our discussion on the important physical  
consequences of our interpretation, leaving
general results of our investigation to be published elsewhere.

The PSM has been successfully applied to normally deformed nuclei \cite{psm}
as well as SD nuclei
in various mass regions \cite{sd190a,sd130a}.
For details of the PSM theory we refer to the review article of
Hara and Sun \cite{psm} and to the published computer code
\cite{psm_code}.
In the PSM, the many-body wavefunction is a superposition of
(angular momentum) projected multi-quasiparticle states,
\begin{equation}
| \psi^I_M \rangle ~=~
\sum_{\kappa} f_{\kappa} \hat P^I_{MK_\kappa}
| \varphi_{\kappa} \rangle ,
\label{ansatz}
\end{equation}
where $| \varphi_{\kappa} \rangle$ denotes basis states consisting of
the quasiparticle (qp) vacuum, two quasi-neutron and -proton, and
four qp states for even-even nuclei.
The dimension of the qp basis
in the present calculation is about 50.
Since $^{60}$Zn has a deformation of $\beta_2 = 0.47$ \cite{zn60},
the deformation of our basis is
fixed at
$\epsilon_2 = 0.45$ for all
nuclei calculated in this paper.
Three full major shells ($N =$ 2, 3 and 4)
are employed for neutrons and for protons
(with a  frozen $^{16}$O core).
For the Nilsson parameters $\kappa$ and $\mu$ we take the
values of Ref.\ \cite{tod}.
Two-body interactions are then diagonalized 
in the basis generated using the above deformed
mean field with angular momentum projection.  

We use the usual separable-force Hamiltonian \cite{psm}
\begin{equation}
\hat{H}= \hat{H_0}-\frac{\chi}{2} \sum_ {\mu} \hat{Q}^+_{\mu}
\hat{Q}_{\mu}-G_M \hat{P}^+ \hat{P} - G_{Q} \hat{P}^+_{\mu} \hat{P}_{\mu}
\label{ham}
\end{equation}
with
spherical single-particle,
residual quadrupole--quadrupole, monopole pairing, and quadrupole
pairing terms.  The strength $\chi$ of the quadrupole--quadrupole term
is fixed self-consistently with the deformation,
so it is not a true parameter 
\cite{jorge}.
Lack of SD data precludes determining the pairing
interaction strength 
from experimental odd--even mass differences in a systematic way,
so we have used the prescription introduced in
Ref.\ \cite{sd130a}, which corresponds in this case to
multiplying the monopole pairing strengths $G_M$ of
Ref.\ \cite{tod} by  0.90 to accommodate the relative increase in the
size of the basis for the present calculation.
This amount of reduction is consistent with the
principles described in Ref.\ \cite{Szym.61}.
For the quadrupole pairing interaction $G_Q$, a ratio $C =
G_Q/G_M$ = 0.28 is used, the same value used in the heavy SD nuclei
\cite{sd190a}.

To illustrate the physics,
the calculated band diagram (energy for the projected
basis states in Eq.\ (\ref{ansatz})
as a function of spin; see Ref.\ \cite{psm} for a further interpretation of this
diagram)
is shown in Fig.\ 1 for $^{60}$Zn.
The 2-qp states
correspond to the group of bands starting around 4--5 MeV in energy
(solid lines for neutrons and dotted lines for
protons).  Among these bands, we observe that
two behave in a unique
way: at the bandhead they lie a little higher than the other
2-qp bands, but rapidly decrease relative to the other bands
as the system rotates.  Thus, in the initial band crossing region
these two bands are on average 2 MeV lower
than the other 2-qp bands.
The 2-qp states exhibiting this behavior correspond to
the neutron and the proton 2-qp state  
coupled from $K = {1\over 2}$ and $K = {3\over 2}$ particles
in the $g_{9/2}$ orbital to a
total $K = 1$.
The band corresponding to
the $g_{9/2}$ proton 2-qp states crosses the ground band
(the first band crossing)
and becomes the lowest band beyond $I = 14$.

A group of 4-qp states is illustrated in
Fig.\ 1 as the set of dashed lines starting around  8--9 MeV.
One of these that is flat in low spin region
is constructed from the above-mentioned two
$g_{9/2}$ pairs of neutrons and protons that are the most favorable
2-qp states in energy.
The $g_{9/2}$ proton 2-qp band
is crossed between
$I = 18$ and 20 by this 4-qp state
(the second band crossing).
Thus, the important multi-quasiparticle states that lie lowest in energy
for the spin range to be considered
are composed entirely from $g_{9/2}$ orbitals
and we may expect that quasiparticle states from other orbitals
(e.g., $f_{7/2}$ or $p_{3/2}$)
will play a less important role near the yrast line.

From the preceding discussion, we conclude that
high spin physics near the yrast line
in the SD even--even, mass-60 nuclei should be governed by
crossings and interactions between bands built upon
neutron and proton $g_{9/2}$ quasiparticles.
Because the single-particle state density is low, 
we may further expect
the influence of
band crossings and interactions to fluctuate drastically
along isotopic chains.
On the other hand, states built upon
quasiparticles from other orbitals occur at much higher energies. They
can contribute to the collective
quantities (e.g. the collective portion of the angular momentum
and the
total electric quadrupole moment), but not strongly to
quantities dominated by the quasiparticle properties.

We show the calculated energy spectra in terms of the
transitional energy $E_\gamma$ in Fig.\ 2 and dynamical
moments of inertia $\Im^{(2)}$ in Fig.\ 3 for the even--even isotopic chain
$^{60 - 66}$Zn.
Comparisons with experimental data are shown
where data are available.
For $^{60}$Zn, both $E_\gamma$ and $\Im^{(2)}$ agree with
data reasonably well 
(data has the peak at $I = 20$, while the
calculated one is at $I = 18$).
The SD band in $^{60}$Zn is linked experimentally to
the known low-lying states \cite{zn60}, 
so the spin of this band
is known. Thus this agreement supports the choices of 
interaction strengths used in the present
calculation.
We can then predict spins for other SD bands where no linking
transitions are observed. For $^{62}$Zn, the best agreement
between theory and data 
corresponds to placing the measured first SD $E_\gamma$
at $I = 20$ (see, Fig. 2),
thus predicting this transition to be from the state $I = 20$
to $I = 18$. This agrees with
the assignment proposed previously by Afanasjev {\it et al} \cite{ring}.

In Fig.\ 3,
there are in general two peaks in the $\Im^{(2)}$ plots that
reflect the two successive band crossings discussed above.
The first occurs at $I \approx 12$, with the location
and size being similar for each of the 4 isotopes. This is because
the first crossing is mainly the $g_{9/2}$ proton pair crossing, which
is relatively constant within this isotopic chain.
However, the next band crossing, caused by a 4-qp state
of $g_{9/2}$ neutron and proton pairs, leads to
very different consequences for each individual SD band;
this implies
significant theoretical constraints for the possibility of observation,
as we now discuss.

The Projected Shell Model is known to give a good description of band
crossings in
heavier nuclei where it has been tested extensively (for example,
see
Ref.\ \cite{ta175}).
The nature of the crossing (e.g., whether the peak in
$\Im^{(2)}$ is sharp or gentle) is related to the 
angle between crossing bands \cite{psm}. A small angle
spreads the interaction over a wide angular momentum
range, thus producing a smoother change. A large angle
implies that the bands interact over a narrow
angular momentum range and a sudden discontinuity can occur.
In our case, a smaller crossing angle is seen just before $I = 20$
for $^{60}$Zn, producing a smoothed interaction (see Fig.\ 1). In fact, the
two peaks in $\Im{(2)}$ caused by first and second band
crossings merge in this case, resulting in one wide
and smooth peak ranging from low to high spins. However, a larger
crossing angle is found at $I = 20$ for $^{62}$Zn.
Thus, in the $\Im{(2)}$ plot for $^{62}$Zn a clear separation
of the two peaks is seen, with the second one at $I = 20$
being much higher. If this discontinuity is pronounced, it
may be expected to set a lower
limit in angular momentum
for observation of such a SD band with weak intensity,
while the upper limit is determined by the band termination spin
\cite{ring}.
This explains succinctly why the observed SD band in $^{60}$Zn is
long, while in the neighboring $^{62}$Zn,
where one might naively expect similar behavior, the observed band
is very
short.

Galindo--Uribarri {\it et al.} reported a rotational SD band in $^{64}$Zn 
\cite{zn64}. Because of the strong dipole
transitions discovered in their work, this band appears not to belong
to the same type of bands (SD yrast bands characterized by
even integer spins only) discussed above
\cite{commu}. An important question
is why the usual SD yrast band has not been seen in $^{64}$Zn.
We find that, due to different neutron shell
fillings, the position of the $g_{9/2}$ neutron 2-qp band
is shifted higher in energy for this case, which in turn pushes the 4-qp band
higher. Consequently, the second band crossing spin is shifted
to $I = 22$, a spin which is even closer to the band
termination. In addition, this second band crossing is very
sharp (see Fig.\ 3). If 
the experimental analysis were not able to follow
the population over the sharp second band crossing,
there would be at most three or four transitional gamma-rays
to measure, making observation of the SD yrast band in
this nucleus difficult.

Going to the next isotope, $^{66}$Zn, a different
picture appears. Because of the shift in neutron Fermi level,
the pair of $g_{9/2}$ neutrons
contributing to the 4-qp state is changed from $K = {1\over 2}$ and
$K = {3\over 2}$ particles to $K = {3\over 2}$ and $K = {5\over 2}$
(still coupled to total $K = 1$).
Because of the even higher energy and steeper curvature
of this 4-qp band, we find that
it crosses the proton 2-qp band
at $I = 26$ at a very small angle.
In fact, one can hardly see in the $\Im^{(2)}$ plot
that there is a band crossing.
Thus, our calculation suggests that there should be a much better
chance to observe a long SD yrast band in $^{66}$Zn, where
no experiment
has yet been reported \cite{commu}.

It has been demonstrated previously \cite{ta175} that the position of band
crossings can be shifted systematically to
higher spin by a stronger quadrupole
pairing interaction. Therefore,
the discrepancy mentioned in the $\Im{(2)}$ plot in the $^{60}$Zn calculation
(the theoretical peak
occurs two spin units too early) could be improved if a larger
quadrupole pairing interaction were employed. We have not
introduced this refinement
because for this particular $N = Z$ nucleus we may expect that
neutron--proton pairing correlations may also play a
role. For example, Ref. \cite{wyss} found that the
$T = 0$ pairing becomes significant at very high spins 
( where the g9/2 orbital became important)
in the lighter $N = Z$
nucleus $^{48}$Cr. 
This neutron--proton pairing has not been included in the
present calculation or in the calculations of Ref.\ \cite{cu58,ring}.
Explicitly including
the p-n pairing in the PSM is of interest for future work.

When we calculate the pairing gaps using the 
total many-body wavefunction, 
we find that
both neutron and proton pairing is significant
at $I = 0$ (gaps of about 0.9 MeV). However,
there is a rapid drop in pairing gaps near the
first band crossing.
Beyond $I = 18$,
they assume small, nearly constant values
corresponding to about 40$\%$ of their initial values.
All measured SD bands in the mass-60 region
are in the spin range beyond $I = 18$.
Thus, our results may provide an understanding of 
the success of mean-field
calculations, all of which have neglected pairing correlations completely
\cite{cu58,ring}.
Details will be reported in a forthcoming paper.

To summarize,  the Projected Shell Model has been used to carry out
the first study of SD mass-60 even--even nuclei using techniques
that go beyond the mean field.
In contrast to the heaviest SD 
systems where coherent motion
of many nucleons is important and alignment in specific orbits is less
significant,
it is found that alignment
of $g_{9/2}$ proton
and neutron pairs dominates the 
high-spin behavior in these lightest SD nuclei.
Because of this, and the low level densities expected
for this mass region near the Fermi surface,
we find that the nature of the SD bands can fluctuate strongly
with shell filling in even--even isotopic sequences.
Calculations for the even--even Zn isotopic chain provide an explanation for
the bands already observed, and make specific predictions about which nuclei
are best candidates for long rotational SD sequences in this region.
Because our calculations go beyond the mean field, they can be used to
check various assumptions of the mean field descriptions.  For example, we
have calculated the pairing gaps dynamically and find that they  
generally are not small at low spins, but drop rapidly to non-zero
 but relatively
small values in the region where data
are available, thus providing a partial
microscopic justification for the uniform
neglect of pairing in all mean field calculations reported to date.
Finally, for the only case in this region where the spin has been measured, 
our calculated spin agrees with the measured spin without parameter
adjustment.  This, coupled with numerous previous correct predictions of spin
for SD bands in the mass-130 and mass-190 nuclei permits us to
predict
theoretical spins with confidence for those cases where they have not yet been
measured.  

Valuable discussions with A. Galindo--Uribarri, P. Ring, A.V. Afanasjev
and D.J. Hartley are acknowledged.
One of us (J.-y. Z.) is supported by the U.~S. Department
                            of Energy through Contract No.\
                            DE--FG05--96ER40983.

\baselineskip = 14pt
\bibliographystyle{unsrt}

\begin{figure}
\caption{ 
Band diagram calculated for SD $^{60}$Zn. Black dots are the
yrast states after band mixing at each spin, which are used
to plot the theoretical curves in Figs. 2 and 3. 
}
\label{figure.1}
\end{figure}

\begin{figure}
\caption{ 
PSM results for $E_\gamma (I) = E(I) - E(I-2)$ in SD yrast bands, 
and comparison with experiment where data are available 
(Ref. (\protect\cite{zn60}) for $^{60}$Zn 
and Ref. (\protect\cite{zn62}) for $^{62}$Zn).
}
\label{figure.2}
\end{figure}

\begin{figure}
\caption{ 
PSM results for $\Im^{(2)} (I) = 4 / (E_\gamma (I) - E_\gamma (I-2))$ 
in SD yrast bands, 
and comparison with experiment where data are available. 
}
\label{figure.3}
\end{figure}

\end{document}